# In-situ total scattering investigation of crystalline ordering in amorphous ion-beam sputtered thin films for interferometric gravitational wave detectors

Alberto Martinelli
Giulio Favaro
Giacomo Ciani
Livia Conti
Jean-Pierre Zendri
David Hofman
Massimo Granata
Angela Trapananti
Flavio Travasso
Laura Silenzi
Valeria Milotti
Marco Bazzan

**Abstract**

Amorphous tantala (a-) is an important optical material used in a number of high-precision optical applications, including gravitational wave interferometry. In this paper, we study *in-situ* the structural changes that occur in amorphous ion-beam sputtered coatings during an annealing treatment by means of a synchrotron radiation scattering experiment . The scattering signal is measured as a function of time on a large range of the Q-space. X-Ray diffraction and Rietveld analysis are used to study crystallization during the annealing treatment, whereas pair distribution function (PDF) analysis allows to inspect the structural changes occurring during the amorphous–to-crystalline transition. Our findings indicate that several structural rearrangements occur in parallel, namely a first quick establishment of a “backbone” structure in the cationic substructure appearing on a rather extended range (up to 100 Å), followed by a progressive rearrangement of the oxygen atoms environment which gradually increases the crystallinity of the structure.

## 1 Introduction

In 2015 the two LIGO detectors successfully detected the first gravitational wave (GW) signal (LIGO and al 2016). A few years later, the Virgo interferometer joined the observation run, significantly increasing the location accuracy and the detection rate (al. 2023). In the corresponding observation run, the merger of a binary neutron star system was detected and triggered the observation of the same event in a number of different observatories and with different signals, marking the birth of multi-messenger astronomy (LIGO and al 2017). These are just a few of the incredible scientific achievements made possible by the development of gravitational wave detectors.

These instruments, among the most sensitive detectors ever built, are essentially large-scale laser interferometers sensing incredibly small relative displacements of their mirrors (in this context called test masses) of the order of $10^{-21}$, upon the passage of a GW (Saulson 2017). The test masses are large silica cylinders suspended inside a large vacuum system by means of a chain of mechanical super-attenuators to decouple the mass motion from the environment (al. 2016; al. 2014). In order to achieve the high reflectivity needed to work as mirrors, the test masses are coated with a dielectric Bragg multilayer made up of a stack of highly transparent optical layers with different refractive indexes (M. Granata et al. 2020b). Due to the extreme performances targeted for GW detection, the mirrors must satisfy very stringent requirements in terms of low absorption and low scattering (below 1 and 10 ppm, respectively) and low transmittance (10 ppm) (Degallaix et al. 2019; Pinard et al. 2017). Furthermore, thermally activated fluctuations of the coating's atomic structure are an important source of noise limiting the interferometer performances (Harry et al. 2006). Because of the fluctuation-dissipation theorem (Callen and Greene 1952), multilayer stacks with low mechanical dissipation are the ones performing best in this respect (McGhee et al. 2023; Vajente et al. 2021; M. Granata et al. 2020a). For this reason, together with extreme optical performances, coating materials must possess very specific mechanical properties as well.

Amorphous tantalum oxide (a-), also known as tantala, is a well-known optical material exploited in the fabrication of Bragg multilayers, together with amorphous silicon dioxide, or silica (a-). Their refractive indexes are respectively 2.05 and 1.45. This provides a sufficient optical contrast to realize an efficient mirror at 1064 nm which is the wavelength employed in present GW detectors. Amorphous /: multilayers with a quality suitable for GW detectors are fabricated by Ion Beam Sputtering deposition (IBS) at Laboratoire des Matériaux Avancés (LMA) in Lyon, France (*Lma.in2p3.fr*, n.d.). As a standard step of the production process, the layers are annealed in air for 10 hours at 500°C to help release mechanical stresses induced by the deposition process and improve the optical and mechanical properties of the coatings as their structure is allowed to attain a more relaxed, energetically favorable state (Anghinolfi et al. 2013). Such improvements are associated with the development of a medium range reordering of the amorphous structure (Hart et al. 2016; Amato et al. 2020; Prasai et al. 2019, 2023; Shyam et al. 2016). In a previous work (Martinelli et al. 2021) it has been shown that the monoclinic *C*2 structural model of the Z- (Zibro et al. 2016) phase provides a good starting structural model for fitting the PDF data collected for a- thin films. At the same time, an EXAFS study (Bassiri et al. 2015) reported that at the local level, the Ta atoms of the amorphous tantala coatings deposited by IBS appear to be less coordinated with oxygen than what would be expected from most crystalline structures and that no significant structural changes are observed on the short range with respect to the as-deposited sample, if the sample is annealed at moderate temperatures and/or durations.

It is important to notice that standard annealing recipes are determined somehow empirically so that, at present, there is not a fundamental criterion to establish an "optimum" engineered annealing treatment. For instance, performing a longer annealing could bring additional improvements to the multilayer. However, if the annealing is performed for too long or at too high temperatures, the amorphous coatings start to form crystalline micro-regions (Favaro et al 2024), which are detrimental to the optical quality of the coatings. In particular, in silica/tantala multilayers, it is the latter showing a more pronounced tendency to crystallize at lower temperatures, constituting the limiting issue in the development of a possibly more effective annealing treatment.

In an effort to better understand the changes promoted by thermal treatments and to develop more effective post-deposition treatments, it is of great importance to follow in detail the whole process of structural rearrangement occurring during an annealing process, from the beginning of the treatment to the onset of the crystallization. Furthermore, as many other oxides, tantala can crystallize in a large variety of different polymorphs which may possess different optical and mechanical properties. It is therefore of interest to elucidate which structure is formed as a consequence of the crystallization process. Also the microstructure of the newly-formed grains is of great importance as their size distribution crucially affects the scattering properties of the annealed film.

In this work we address this task by an in-situ total x-ray elastic scattering investigation with synchrotron radiation. IBS-deposited coating samples are studied in-situ by collecting x-ray scattering data while the samples are annealed. The data are recorded as a function of time at fixed temperature, allowing us to precisely investigate the onset and the evolution of the amorphous to crystalline transition. In the total scattering experiment, long range ordering is studied by means of X-ray diffraction analysis, while short range ordering is accessed via measurement of the Pair distribution Function (PDF). Our results allow tracking the structural evolution of the crystalline regions inside the films during the annealing treatment.

*Evolution of the diffraction patterns as a function of time during isothermal annealing at 690 °C for thin films of amorphous a- characterized by different thickness. On the top all the diffractogram for each of three analyzed sample are presented. On the bottom three different diffraction figures collected at different times on sample TF05 (left: start of the process, center: after one hour of treatment, right: at the end of the process) showing how during the heating treatment the crystallization produces the appearance of the Laue diffraction rings.*

# 2 Experimental Methods

## 2.1 Samples

Thin films of amorphous were deposited by ion-beam sputtering on (100) single-crystal Si substrates (75 mm diameter, 450 $\mu$m thick, p-doped to 0.1-1 Ω.cm) in a Veeco SPECTOR system, using argon ions as sputtering particles. The ion beam voltage and current were 1.3 kV and 0.6 A, respectively. The base pressure inside the deposition chamber was below $10^{-5}$ mbar before deposition. During the process, a flow rate of 18 standard cubic centimeters per minute (sccm) of argon was injected into the ion source while a flow of 25 sccm of oxygen was injected into the coating chamber, resulting in a working pressure of the order of $10^{-4}$ mbar. Three different samples were analyzed, characterized by different film thicknesses, namely 0.5 $\mu$m, 1.0 $\mu$m and 2.5 $\mu$m (hereinafter, these samples are referred to as TF05, TF10, and TF25, respectively).

## 2.2 Data Acquisition

To investigate the crystallization process, *in-situ* data collection was carried out during thermal treatment at 690 °C. This temperature was selected, after a preliminary laboratory investigation (not shown in this work) as a compromise between a good time resolution of the experiment and a sufficiently fast evolution of the crystallization process to be able to measure it within the allocated beamtime. The preliminary investigation has been carried out with a Panalytical MRD X'pert X-ray diffractometer equipped with an Anton-Paar AP-900 thermal stage that allowed to determine the onset of crystallization and the time necessary to track its full evolution as a function of temperature. This study showed also that within the 500°C-800°C temperature range, the X-ray spectrum of the crystallized phase remains invariably the same, indicating that changing the annealing temperature modifies markedly the velocity of the crystallization but not the final crystal phase.

Elastic X-ray scattering data were collected at the ID11 beamline of the European Synchrotron Radiation Facility (ESRF-Grenoble, France) using a wavelength $\lambda$ = 0.1582 Å and a grazing-incidence geometry, by means of which only the a- is illuminated by the X-ray beam (thus avoiding the signal of the Si substrate).
The diffraction pattern has been measured using a FReLoN camera suitable for 2D acquisition. This is optimized for very rapid readout, allowing full (2048 x 2048 pixels, 16 bit) frames to be read out in 240 ms, which can be further increased binning or by the use of regions of interest.(al. 2010)

## 2.3 Data analysis

For the crystalline phases, structural refinements were carried out using X-ray powder diffraction (XRPD) data applying the Rietveld method (Young 1993) and using the FullProf Suite (Rodríguez-Carvajal 1993); in particular, a file describing the instrumental resolution function (obtained by using a standard sample) and a Thompson-Cox-Hastings pseudo-Voigt convoluted with an axial divergence asymmetry function were used during calculations.
In the structural model, the cationic site occupations were fixed to the nominal compositions. Given that all the atomic site coordinates are constrained by symmetry, in the final cycle the following parameters were refined: the scale factor; the zero point of the detector; the background; the unit cell parameters; the isotropic displacement parameters. Parameters defining line profiles provide information on the size of diffracting domains and on structural deformation (microstrain); for this scope, the anisotropic strain parameters were also included in the refinement.
To gain information on the local and medium range structure, Pair Distribution Function (PDF) analysis (Billinge and Thorpe 1998) was carried out. Reduction of total scattering data was achieved using PDFgetX3 software (Juhás et al. 2013) with a maximum scattering vector modulus $Q_{max} = 14.6$ Å. In this way, the following functions were obtained: $I(Q)$ (i.e. the background-corrected diffraction intensity), $S(Q)$ (the total-scattering structure function with intensities normalized by average scattering factors and corrected by a polynomial fit) and $G(r)$ (the reduced PDF) (Egami and Billinge 2003).
Full-profile fitting of the $G(r)$ function was carried out using the PDFgui software (Farrow et al 2007). A standard sample was analysed under the same experimental set-up to describe the experimental resolution effects (Qdamp and Qbroad parameters); these parameters were then fixed during the fitting of the $G(r)$ functions obtained for .
In the last cycle of the fitting, the following parameters were refined: the scale factor; the unit-cell parameters; the atomic positions not constrained by symmetry; the isotropic atomic displacement parameters (distinct parameters for distinct atomic species); the dynamic correlation factor; a parameter including scatterer size effects in the PDF (spdiameter). PDF data fitting was carried out by adopting different length scales, to unveil the evolution of structural properties as a function of distance.

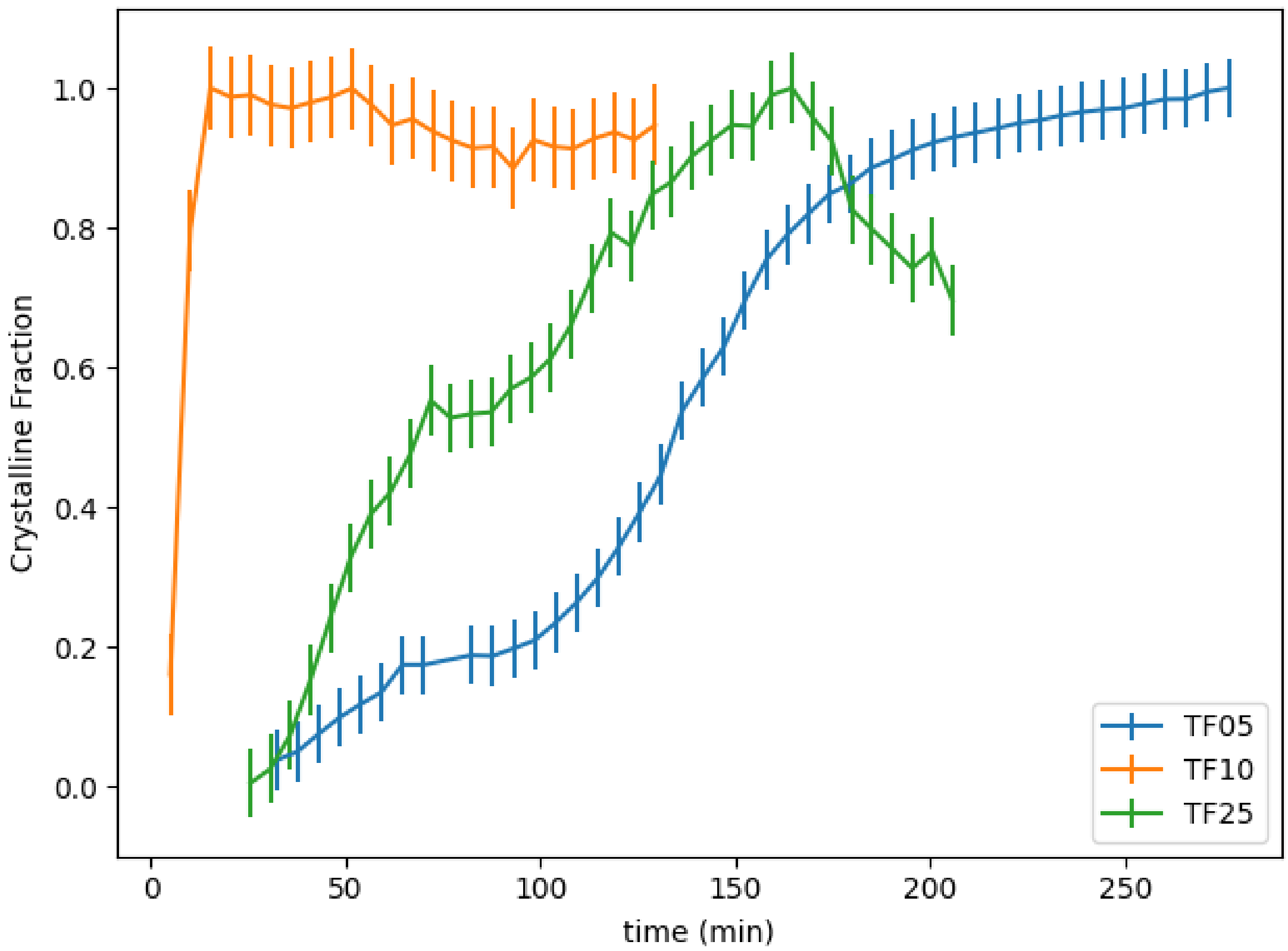

*Estimate of the main crystalline phase volume fraction in each of the three samples as a function of the annealing time*

# 3 Results and Discussion

## 3.1 X-ray diffraction

Figure 1 shows the evolution of the XRPD patterns during thermal treatments at 690 °C for the 3 thin films of amorphous a- characterized by different thicknesses. At the beginning of the experiment, all the samples were in their amorphous state, as evidenced by the absence of any sharp diffraction peak. As the time goes on, the formation of crystalline regions becomes apparent with the emergence of diffraction peaks from the amorphous background. This process can be better appreciated in Fig. 2 where an estimate of the crystalline fraction as a function of time for the three samples is provided. The estimate has been carried on by normalizing the area of the most intense diffraction peak with its maximum value across the process assuming that the samples reach a fully crystallized state. In all samples we observe that the nucleation and growth of crystalline regions does not start as soon as the annealing plateau is reached, but occurs after an "induction time" varying from sample to sample and of the order of several tens of minutes. This is related to the transient period preceding the formation of crystalline nuclei and their subsequent growth, in accordance with classical nucleation theory (see e.g. (Schmelzer 1995)). The different induction times observed from sample to sample may be due to the fact that different degrees of relaxation of the amorphous films decisively influence the onset of the crystallization process (Abyzov et al. 2023). In particular, sample TF05 shows a somewhat slower evolution with respect to TF25 and especially to TF10, which appears to be the first one to crystallize.

*Le Bail fitting plot for samples TF10 (left) and TF25 (right). XRPD data collected at the end of the thermal treatment.*

| Sample | Space Group | Cell Parameters (Å, deg) | | | | | | Size (Å) | R-factors (%) | |
|---|---|---|---|---|---|---|---|---|---|---|
| 3-8 | | a | b | c | $\alpha$ | $\beta$ | $\gamma$ | | $R_{Bragg}$ | $R_p$ |
| TF05 | P2/m | $6.2721(6)$ | $4.0619(1)$ | $5.2009(1)$ | 90 | $93.09(1)$ | 90 | $\sim 105$ | 0.767 | 1.15 |
| TF10 | P2/m | $6.278(1)$ | $4.046(1)$ | $5.3177(1)$ | 90 | $92.98(1)$ | 90 | $\sim 75$ | 1.01 | 2.34 |
| TF25 | P6/mmm | $3.6236(1)$ | $3.6236(1)$ | $11.6477(1)$ | 90 | 90 | 120 | $\sim 60$ | 0.046 | 0.054 |

It is evident that at the end of the thermal treatment the samples are characterized by very similar diffraction patterns, but a closer inspection reveals that both TF05 and TF10 samples exhibit a diffraction peak at ~2.2 $Å^{-1}$ that is absent for TF25. It is interesting to observe that this peak at ~2.2 $Å^{-1}$ is initially absent in the TF05 sample and it is actually formed in a later step of the crystallization process; conversely, this same peak is observed in the very early steps of the crystallization of the TF10 sample. This indicates that slightly different structural modifications may compete during annealing.

The XRPD patterns collected at the end of the thermal treatment (high degree of crystallinity) were indexed using the DicVol program included in the FullProf Suite; in this way, a set of possible space groups was selected and then the whole diffraction pattern profile fitting was carried out applying the Le Bail method. An excellent fitting of the TF25 XRPD pattern was achieved using the *P*6/*mmm* space group, whereas the patterns of the TF05 and TF10 samples are better fitted with a *P*2/*m* space group (Figure 3 and Table [tab:XRPD]). Despite the large number of polymorphic modifications reported for stoichiometric , to our knowledge none of them matches the structures listed in Table [tab:XRPD]. It is however worth mentioning that Khitrova and Klechkovskaya (Khitrova and Klechkovskaya 1980) reported the formation of a polymorphic modification of in thin films crystallizing in the *P*6/*mmm* space group with a *c*-axis value (11.52 Å), similar to that obtained in our sample by Le Bail fit, and a doubled value of the *a*-axis (7.16 Å).

Remarkably, a group-subgroup relationship holds for the space groups *P*6/*mmm* and *P*2/*m* with lattice parameters listed in Table [tab:XRPD] , consistently with the experimental observation that the peak at ~2.2 $Å^{-1}$ is initially absent in the TF05 sample, but grows in the final stage of the thermal treatment. In other words, a *P*6/*mmm* → *P*2/*m* desymmetrization can comply with the structural change evidenced by the appearance of the peak at ~2.2 $Å^{-1}$ in the XRPD data of the TF05 sample.
It is interesting to observe that the lattice parameters of the TF25 sample (Table [tab:XRPD]) are quite similar to those of the hexagonal polymorph reported by Yamaguchi et al. (Yamaguchi et al. 1988), crystallizing with structure type, but with a tripled *c*-axis. Indeed, a 3*c* structural model can be derived from structural data reported in ref. (Yamaguchi et al. 1988) and used for Rietveld refinement, providing a reasonably acceptable fitting ($R_{Bragg}$ = 17.8 %; Figure 4).

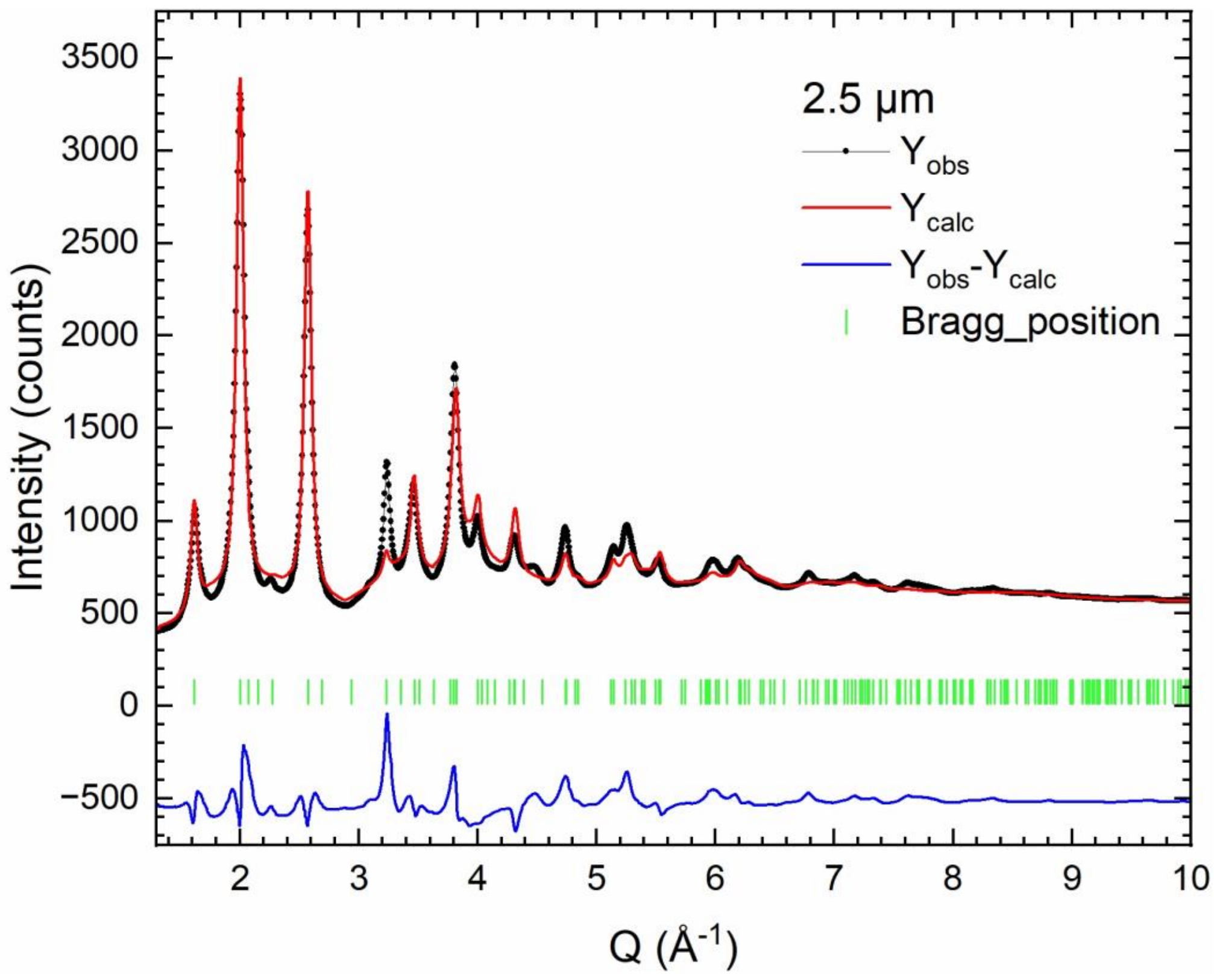

*Rietveld refinement plot obtained using a P6/mmm structural model derived from that reported in ref (Bassiri et al. 2015) after tripling the c-axis.*

However, this result can be obtained only by allowing the oxygen isotropic displacement parameters to assume very high values, which is unrealistic from a physical point of view. By restricting these parameters to meaningful values, the calculated intensities of the main diffraction lines at ~2.0 $Å^{-1}$ and ~3.8 $Å^{-1}$ fall almost to zero in strong disagreement with our data. It can be thus concluded that the *P6*/*mmm* structural model with tripled *c*-axis provides an approximate description of the crystal structure, but the distribution of the oxygen atoms within the structure is significantly more complicated leading to local symmetry breaking. This aspect will be further discussed in section 3.2.

*Evolution of the cell parameters (upper panels) and coherent diffraction domain size as a function of the annealing time for the monoclinic TF05 and hexagonal TF25 samples (after Le Bail fitting).*

By applying a sequential Le Bail fitting of the diffraction patterns for each time frame, the evolution of the cell parameters and diffraction domain size as a function of the annealing time can be inspected in the three samples (Figure 5). Irregularities observed in the domain-size panels can be ascribed to slight changes in the sample positions during data collection (during annealing, on account of thermal expansion, the samples were periodically re-centered to avoid/minimize the substrate contribution). As anticipated, the time evolution of the sample TF05 appears slower than the one of sample TF25 which reaches a stationary condition already after  50 min. Furthermore, at the end of the thermal treatment, the resulting domain size is found smaller for higher film thickness (Table 1); this suggests that the number of nucleation centers increases with film thickness, thus preventing the development of larger domains.

## 3.2 Pair distribution function

Figure 6 shows the superposition of the experimental reduced pair distribution $G(r)$ functions obtained at the end of the thermal treatments for the three samples. The damping of the functions with distance provides direct measurements of the structural coherence of the inspected material; it appears that the size of the crystalline domains increases with the decrease of the thin film thickness, in fair agreement with the data calculated by XRPD analysis (Table [tab:XRPD]).

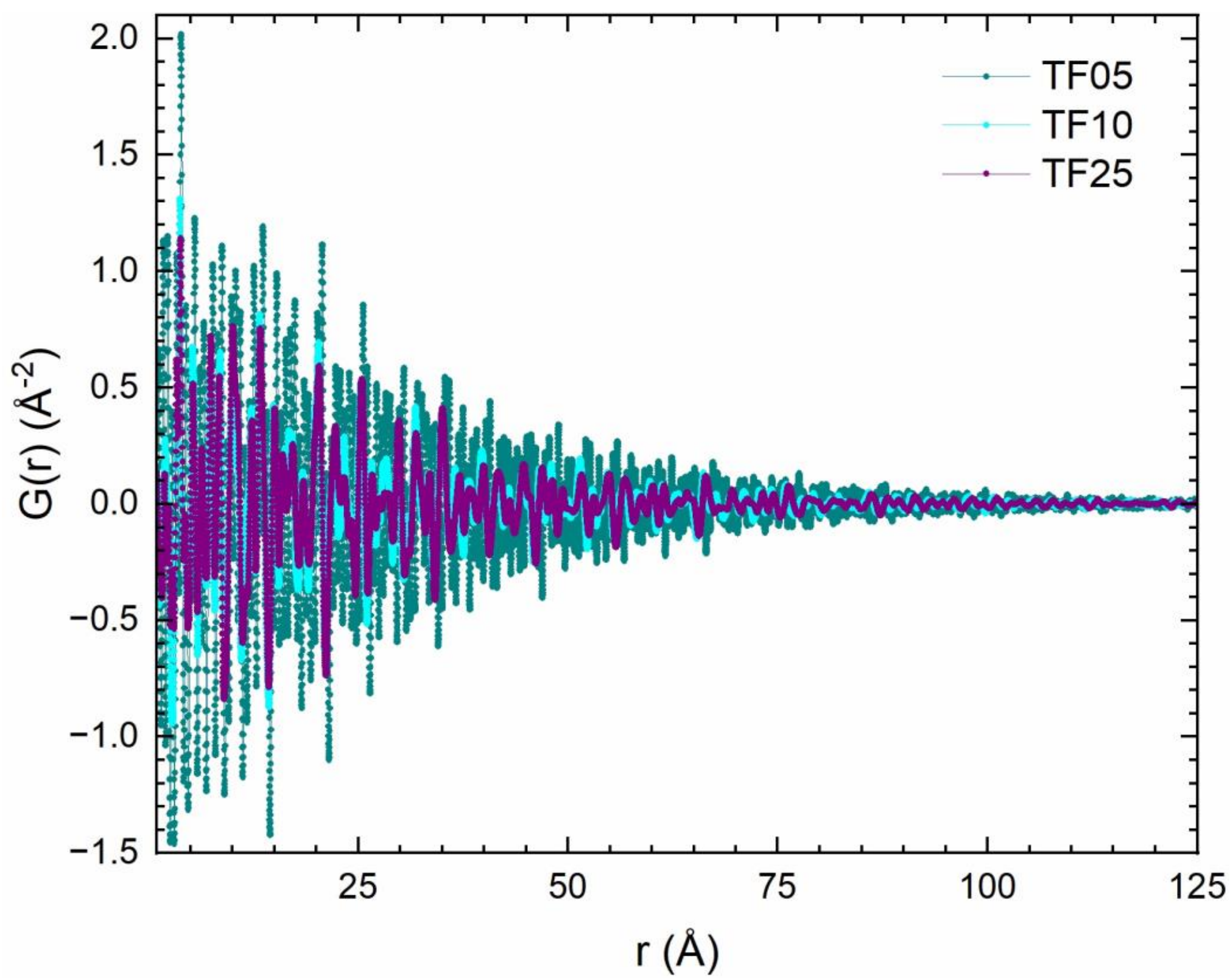

*$G(r)$ functions obtained at the end of the thermal treatments for the three samples. Sample TF25 exhibits a lower degree of structural coherence compared to the others.*

The same 3*c*-*P*6/*mmm* structural model obtained for the TF25 sample and used for the Rietveld refinement was applied to fit the $G(r)$ function obtained at the end of the thermal treatment. On a short/medium-range scale (below ~10 Å) this model is not suitable to describe the observed PDF peaks (Figure 7, on the left; $R_w$ = 40.36%), but in the range between 10 and 35 Å a reasonable good data fitting is obtained (Figure 7, on the center; $R_w$ = 29.34%).

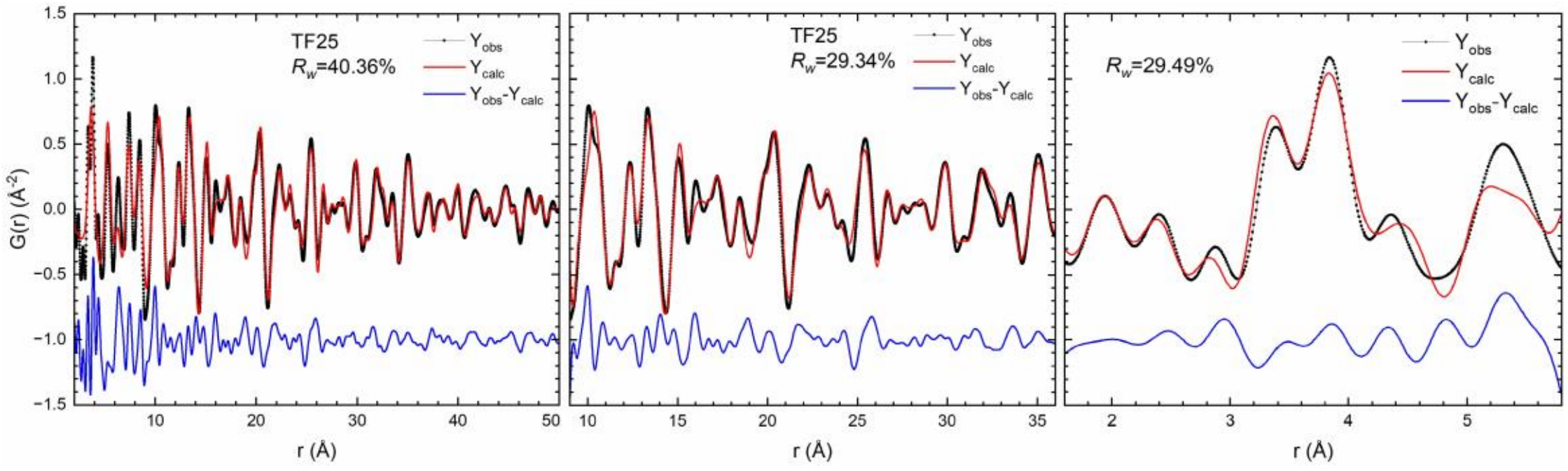

*Experimental $G(r)$ function obtained for the TF25 sample at the end of the thermal treatment. The first two graph show a fit of the $G(r)$ function with a 3c-P6/mmm structural model in the distance range $1.5 < r < 50$ Å and in the distance range $9 < r < 36$ Å. In the last graph the $G(r)$ function has been fitted with a P1 structural model in the distance ranges $1.5 < r < 5.8$ Å.*

The simple structure model (1*c*-*P*6/*mmm* (Yamaguchi et al. 1988)), gives comparable $R_w$ results, in the distance intervals considered (40.17% and 29.35%, respectively). This confirms that the structure type can be taken as the prototypical model to describe the average structure of crystallized , but at the local scale the arrangement of the atoms is more complicated and locally breaks the symmetry, in agreement with the findings of ref. (Martinelli et al. 2021).
A structural model for the local environment complying with the $P2/m$ unit cell obtained via the Le Bail fitting (Table [tab:XRPD]) is unknown. However, in a previous work (Martinelli et al. 2021) it was found that the monoclinic $C2$ structural model of the Z- phase (Martinelli et al. 2021) provides a good starting structural model for fitting the PDF data collected for a- thin films. In particular, acceptable fittings of the $G(r)$ functions were obtained by using a triclinic structural model derived from $P1$ space group. By applying the same $P1$ structural model a reasonably good fitting is obtained on the short range (Figure 7; $R_w$ = 29.49%). Remarkably, this model is able to catch the main features of the $G(r)$ function at the local scale, but fails to provide an acceptable fitting for larger distances (data collected in the 9-36 Å distance range are fitted with $R_w$ =58.12%). This can be interpreted with the fact that averaging on longer distances, the different details of the short-range distortions are masked out to give an overall higher-symmetry structure.

By assuming this structure, specific atomic bonds can be assigned to the peaks of the $G(r)$ function. The peaks at ∼3.4 Å , ∼3.8 Å and ∼4.4 Å are predominantly associated with Ta–Ta atomic correlations, with minor contributions from Ta–O pairs. The features observed at approximately $\sim$ 1.9 Å and $\sim$ 2.9 Å arise primarily from Ta–O correlations, as the contribution from O–O pair correlations is almost negligible.
At this point, it is instructive to inspect in detail the evolution of the $G(r)$ function during thermal treatment; Figure 8 shows selected $G(r)$ functions collected at different times during thermal treatment of the TF25 sample. The first observation is that the short-range structure up to 4.5 Å is already present even before the annealing has started, as expected for an amorphous system. In particular, the peak attributed to the nearest Ta - Ta pairs is well visible in the as-deposited sample (pattern 1). On the contrary, the peak at ∼5.3 Å attributed to the 2nd nearest Ta-Ta pairs, is practically absent, evidencing that structural correlation in the amorphous a- phase is essentially limited to the first coordination sphere. As the time goes by, a long range ordering is established during the annealing. Data in Figure 8 shows that after 50 minutes at our annealing temperature of 680 °C, spatial correlation up to $\sim$ 80 Å is visible. From this moment on, the structural changes are limited to an overall improvement of the crystalline ordering, corresponding to a moderate increase and sharpening of the different peaks.

To better elucidate this phenomenon, we fit the different peaks using a pseudo-Voigt profile and report the relative changes of their height as a function of time (Figure 9). It can be observed that all the peaks within a radius of $\sim$ 4Å remain almost unchanged up to a 10% of their initial height, while peaks at larger distances show a marked increase in the first 50 minutes of annealing, as expected. Still, it is interesting to note that, even if almost all the peaks follow a similar qualitative evolution, the peak at 5.3 Å (corresponding to Ta - Ta bonds in the 2nd shell) starts its evolution immediately after the beginning of the annealing, while peaks at 7.4 Å or 8.5 Å, responsible for Ta - O and O - O bonds in the higher order shells, show a somehow slower evolution. Moreover, the peak at 3.3 Å corresponding to the nearest Ta - Ta pairs ends its evolution faster compared to the Ta - Ta bonds in the 2nd shell. This hints that not all the atomic species follow the same dynamic while the reordering of the structure is taking place. In particular, oxygen atoms seem to rearrange on a different time scale compared to the metallic Ta backbone of the crystalline structure.

Overall, we can interpret our data within the following scheme: when a- is deposited the basic structural building blocks are already present in the form of Ta - O distorted octahedra with a P1 symmetry (mainly caused by the presence of oxygen vacancies). Structural correlation exists up to 4.4 Å. During the annealing, the Ta - O backbone of crystalline is first established up to a length of 80 - 100 Å. However, due to the presence of defects in the oxygen sublattice, the local symmetry of the 3*c*-*P*6/*mmm* is broken even if on average it captures the overall structure of the crystallized phase. As the annealing goes on, oxygen redistributes along the structure, improving the crystallinity of the lattice and partially restoring the symmetry on the short length scales. These two different processes proceed simultaneously; although somehow related, their evolution is not strictly coupled, given that the transition to a crystalline phase, on one hand, and the ordering involving oxygen bonds on the other one, occur with different kinetics during the annealing. Our *in-situ* results complement those reported in ref. (Prasai et al. 2019) where it is indicated that annealing can reduce the number of polyhedra connected through their faces or edges and increases the number of corner-sharing polyhedra.

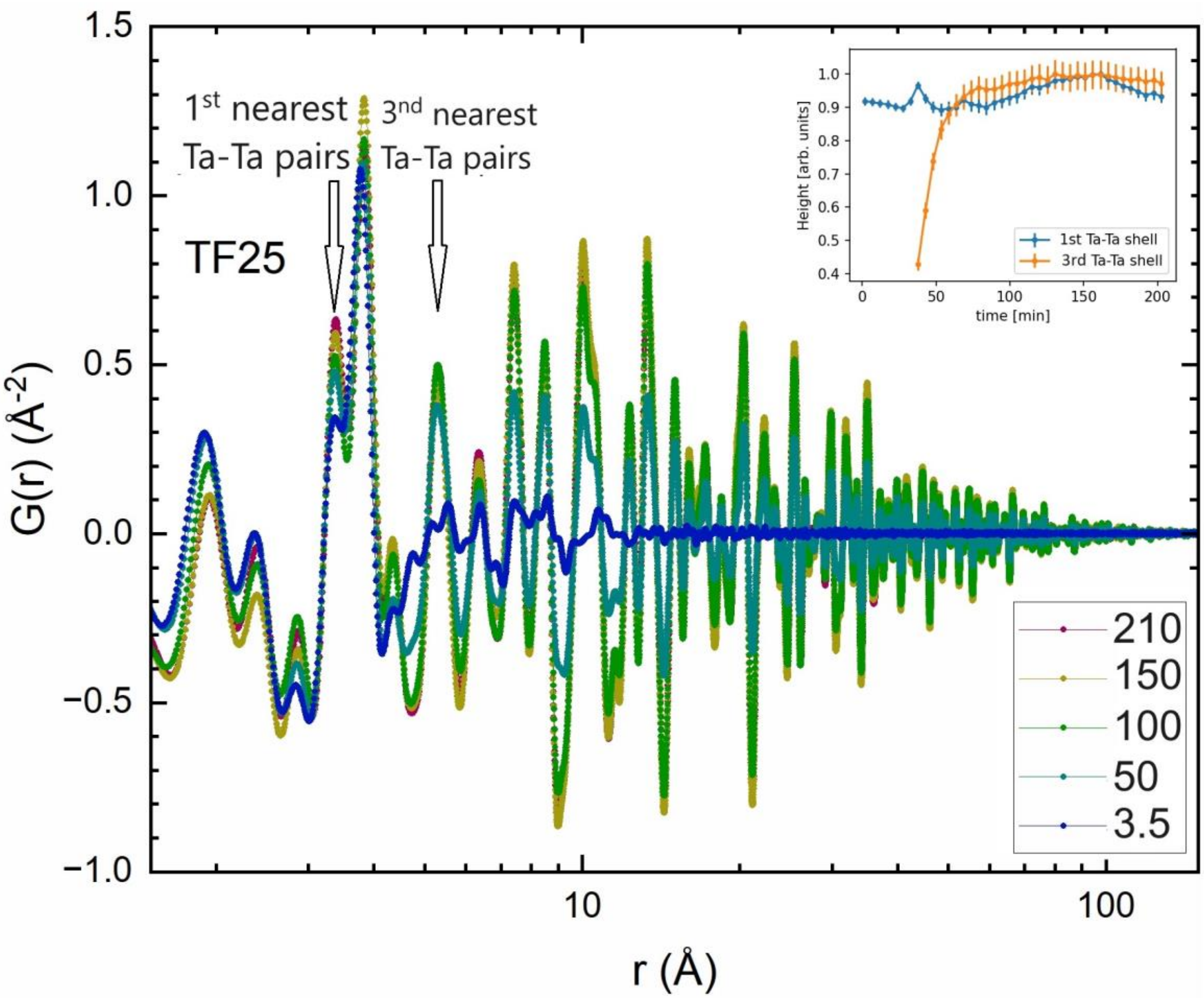

*Selected $G(r)$ functions collected during thermal treatment of the TF25 sample; numbers in the box refer to the acquisition time (min). The inset in the figure shows the height evolution of the first and the third Ta-Ta shells.*

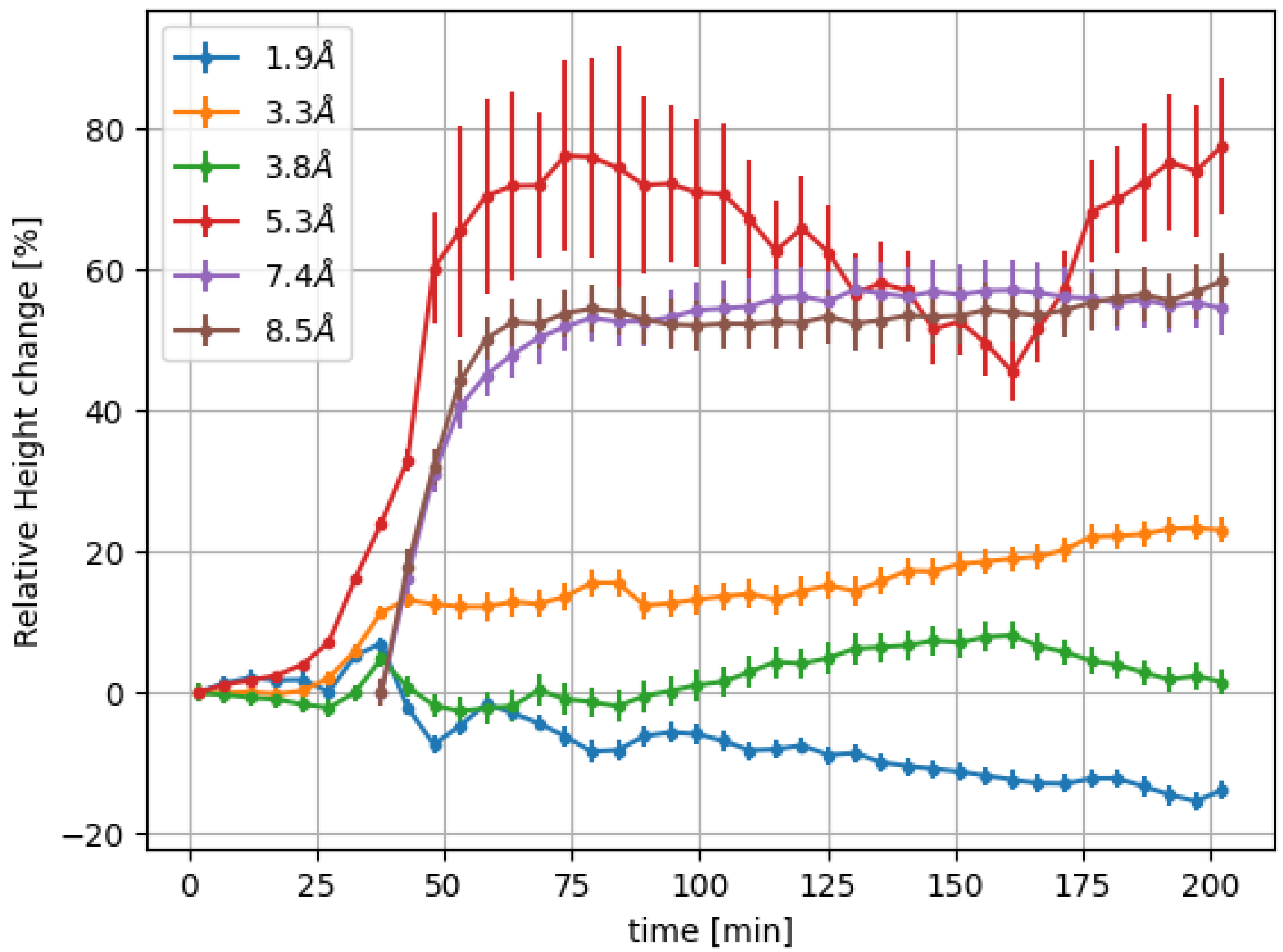

*Comparison of the peak height evolution of a set of the peaks from sample TF25 during thermal treatment.*

# 4 Conclusions

The crystallization of amorphous tantala coatings is studied by means of an *in-situ* X-ray scattering experiment to elucidate the structural changes leading to the onset of the crystalline order. Our data show that crystallization occurs through different parallel processes. Initially the sample is indeed amorphous with the absence of any long-range spatial correlation between the elements. On the local scale, the structure is rather disordered, and can be described in the framework of a triclinic *P*1 structure. When the sample is annealed, as the crystalline fraction gradually increases, two processes take place: (i) the ordering of the basic building blocks of to form the crystalline structure (ii) the re-arrangement of oxygens around the Ta atoms within these blocks. Those two processes, although not completely, occur quite independently.
At the temperatures considered in this experiment, it can be seen that the first rearrangement process of the basic building blocks is rather fast, with spatial correlations up to 100 Å developing in a few tens of minutes after the annealing has begun. Oxygen rearrangement instead takes place with a slower dynamics with a general improvement of the crystalline structure.

The overall structure can be described using *P*6/*mmm* or *P*2/*m* symmetries; indeed, these two space groups are connected by a group-subgroup relation, so that the breaking of the hexagonal symmetry can be seen as a reduction of symmetry on longer distances. Finally, our data shows that the film thickness has an impact on the size of the coherent diffraction domains and on the kinetic of the crystallization process. Samples with smaller thicknesses lead to slower evolution and to the achievement of larger diffraction domains. This can be explained considering that samples with larger thicknesses contain a larger number of nucleation centers, leading to a more finely-grained structure in the fully crystallized stage.

This work was carried out with the support of the European Synchrotron Radiation Facility, (proposal MA-5204; DOI: 10.15151/ESRF-ES-750663650). The authors gratefully acknowledge the support of the Agence Nationale de la Recherche through grant n. ANR-18-CE08-0023 (project ViSIONs), and Dr. Carlotta Giacobbe for her kind assistance during data collection at ID11.